\input harvmac.tex
\input labeldefs.tmp
\writedefs
\overfullrule=0mm
\hfuzz 10pt  
\input epsf.tex
\long\def\ffig#1#2#3{%
\xdef#1{\the\figno}%
\writedef{#1\leftbracket \the\figno}%
\midinsert%
\parindent=0pt\leftskip=1cm\rightskip=1cm\baselineskip=11pt%
\centerline{#3}
\vskip 8pt\ninepoint%
{\bf Fig.\ \the\figno:} #2%
\endinsert%
\goodbreak%
\global\advance\figno by1%
}
\long\def\fig#1#2#3{\ffig{#1}{#2}{\epsfbox{#3}}}
%
%
%
%
\def\frac#1#2{{\scriptstyle{#1 \over #2}}}

\def\dd#1#2{{\partial #1 \over \partial #2}}

\def\({ \left( }\def\[{ \left[ }
\def\){ \right) }\def\]{ \right] }
%

\def\IR{\relax{\rm I\kern-.18em R}}
\font\cmss=cmss10 \font\cmsss=cmss10 at 7pt
\def\IZ{\relax\ifmmode\mathchoice
{\hbox{\cmss Z\kern-.4em Z}}{\hbox{\cmss Z\kern-.4em Z}}
{\lower.9pt\hbox{\cmsss Z\kern-.4em Z}}
{\lower1.2pt\hbox{\cmsss Z\kern-.4em Z}}\else{\cmss Z\kern-.4em Z}\fi}
\def\inbar{\,\vrule height1.5ex width.4pt depth0pt}
\def\IB{\relax{\rm I\kern-.18em B}}
\def\ID{\relax{\rm I\kern-.18em D}}
\def\IE{\relax{\rm I\kern-.18em E}}
\def\IF{\relax{\rm I\kern-.18em F}}
\def\IG{\relax\hbox{$\inbar\kern-.3em{\rm G}$}}
\def\IH{\relax{\rm I\kern-.18em H}}
\def\II{\relax{\rm I\kern-.18em I}}
\def\IK{\relax{\rm I\kern-.18em K}}
\def\IL{\relax{\rm I\kern-.18em L}}
\def\IM{\relax{\rm I\kern-.18em M}}
\def\IN{\relax{\rm I\kern-.18em N}}
\def\IO{\relax\hbox{$\inbar\kern-.3em{\rm O}$}}
\def\IP{\relax{\rm I\kern-.18em P}}
\def\IQ{\relax\hbox{$\inbar\kern-.3em{\rm Q}$}}
\def\IGa{\relax\hbox{${\rm I}\kern-.18em\Gamma$}}
\def\IPi{\relax\hbox{${\rm I}\kern-.18em\Pi$}}
\def\ITh{\relax\hbox{$\inbar\kern-.3em\Theta$}}
\def\IOm{\relax\hbox{$\inbar\kern-3.00pt\Omega$}}


\def\oh{{1\over 2}}


\def\bra{\langle}\def\ket{\rangle}

\def\\#1 {{\tt\char'134#1} }

\catcode`\@=11
\def\Eqalign#1{\null\,\vcenter{\openup\jot\m@th\ialign{
\strut\hfil$\displaystyle{##}$&$\displaystyle{{}##}$\hfil
&&\qquad\strut\hfil$\displaystyle{##}$&$\displaystyle{{}##}$
\hfil\crcr#1\crcr}}\,}   \catcode`\@=12
\def\encadre#1{\vbox{\hrule\hbox{\vrule\kern8pt\vbox{\kern8pt#1\kern8pt}
\kern8pt\vrule}\hrule}}
\def\encadremath#1{\vbox{\hrule\hbox{\vrule\kern8pt\vbox{\kern8pt
\hbox{$\displaystyle #1$}\kern8pt}
\kern8pt\vrule}\hrule}}


\def\d{{\rm d}}
\def\omegasl{\raise.15ex\hbox{/}\kern-.57em\omega}
\def\Lsl{\,\raise.15ex\hbox{/}\mkern-13mu L}
\def\hsl{\raise.15ex\hbox{/}\kern-.57em h}
\def\Hsl{\,\raise.15ex\hbox{/}\mkern-12.5mu H}
\newdimen\xraise\newcount\nraise
\def\xpoint{\hbox{\vrule height .45pt width .45pt}}
\def\udiag#1{\vcenter{\hbox{\hskip.05pt\nraise=0\xraise=0pt
\loop\ifnum\nraise<#1\hskip-.05pt\raise\xraise\xpoint
\advance\nraise by 1\advance\xraise by .4pt\repeat}}}
\def\ddiag#1{\vcenter{\hbox{\hskip.05pt\nraise=0\xraise=0pt
\loop\ifnum\nraise<#1\hskip-.05pt\raise\xraise\xpoint
\advance\nraise by 1\advance\xraise by -.4pt\repeat}}}
\def\Th{Thistlethwaite}

\def\ommit#1{}
%
%
%
\lref\BIPZ{E. Br\'ezin, C. Itzykson, G. Parisi and J.-B. Zuber, 
{\it Comm. Math. Phys.} 59 (1978), 35.}
\lref\KAUF{L.H.~Kauffman, {\sl Knots and physics},
World Scientific Pub Co (1994)}
\lref\BIZ{D. Bessis, C. Itzykson and J.-B. Zuber, 
{\it Adv. Appl. Math.} 1 (1980), 109.}
\lref\DFGZJ{P. Di Francesco, P. Ginsparg and J. Zinn-Justin, 
{\it Phys. Rep.} 254 (1995).}
\lref\HTW{J. Hoste, M. Thistlethwaite and J. Weeks, 
{\it The Mathematical Intelligencer} 20 (1998), 33.}
\lref\MTh{W.W. Menasco and M.B. \Th, 
{\it Bull. Amer. Math. Soc.} 25 (1991), 403; 
{\it Ann. Math.} 138 (1993), 113.}
\lref\PZJ{P.~Zinn-Justin,
{\it Commun. Math. Phys.} 194 (1998), 631.}
\lref\Ro{D. Rolfsen, {\sl Knots and Links}, Publish or Perish, Berkeley 1976.}
\lref\STh{C. Sundberg and M. Thistlethwaite, 
{\it Pac. J. Math.} 182 (1998), 329.}
\lref\Tutte{W.T. Tutte,
{\it Can. J. Math.} 15 (1963), 249.}
\lref\Zv{A. Zvonkin,
{\it Math. Comp. Modelling} 26 (1997), 281.}
\lref\KM{V.A.~Kazakov and A.A.~Migdal,
{\it Nucl. Phys.} B311 (1988), 171.}
\lref\Kaz{V.A.~Kazakov, {\it Phys. Lett.} A119 (1986), 140.}
\lref\KP{V.A.~Kazakov and P.~Zinn-Justin, 
{\it Nucl. Phys.} B546 (1999), 647.}
\lref\PZJZ{P.~Zinn-Justin and J.-B.~Zuber, preprint {\tt math-ph/9904019},
to appear in the proceedings of the 11th 
International Conference on Formal Power Series and Algebraic 
Combinatorics, Barcelona June 1999.}
\lref\DAL{S.~Dalley, {\it Mod. Phys. Lett.} A7 (1992), 1651.}
\lref\HC{Harish~Chandra, {\it Amer. J. Math.} 79 (1957), 87.}
\lref\IZ{C.~Itzykson and J.-B.~Zuber, {\it J. Math. Phys.} 21 (1980), 411.}
\lref\KKSW{V.A.~Kazakov, M.~Staudacher and T.~Wynter,
{\it Commun. Math. Phys.} 177 (1996), 451; 179 (1996), 235;
{\it Nucl. Phys.} B471 (1996), 309\semi
I.~Kostov, M.~Staudacher and T.~Wynter,
{\it Commun. Math. Phys.} 191 (1998), 283.}
\lref\VERSH{A.M.~Vershik and S.V.~Kerov, {\it Soviet. Math. Dokl.}
18 (1977), 527.}
\lref\DK{M.R.~Douglas and V.A.~Kazakov,
{\it Phys. Lett.} B319 (1993), 219.}
\lref\PZJb{P.~Zinn-Justin, preprint {\tt cond-mat/9909250}.}
\lref\KoS{I.K.~Kostov, {\it Mod. Phys. Lett.} A4 (1989), 217\semi
M.~Gaudin and I.K.~Kostov, {\it Phys. Lett.} B220 (1989), 200\semi
I.K.~Kostov and M.~Staudacher, {\it Nucl. Phys.} B384 (1992), 459.}
\lref\DF{P. Di Francesco, O. Golinelli and E. Guitter,
{\it Commun. Math. Phys.} 186 (1997), 1\semi
P. Di Francesco, B. Eynard and E. Guitter,
{\it Nucl. Phys.} B516 (1998), 543.}
\lref\MAK{Y. Makeenko, {\it Nucl. Phys. Proc. Suppl.} 49 (1996) 226.}
\lref\Wit{E. Witten, {\it Surv. in Diff. Geom.} 1 (1991), 243.}
\lref\Kon{M. Kontsevich, {\it Funk. Anal. \& Prilozh.} 25 (1991), 50.}
\lref\IZb{C. Itzykson and J.-B. Zuber, {\it Int. J. Mod. Phys.} A7
vol. 23 (1992), 5661.}
\def\vertex{\epsfxsize=5mm\hbox{\raise -1mm\hbox{\epsfbox{link01.eps}}}}
\def\vertexalt{\epsfxsize=5mm\hbox{\raise -1mm\hbox{\epsfbox{link02.eps}}}}
\def\vertexdbl{\epsfxsize=8mm\hbox{\raise -4mm\hbox{\epsfbox{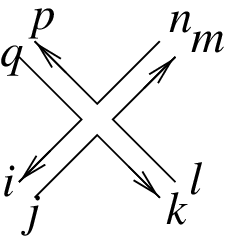}}}}
\def\propagdbl{\epsfxsize=12mm\hbox{\raise -1mm\hbox{\epsfbox{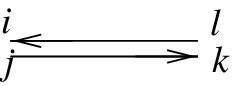}}}}
\def\vertexarr{\epsfxsize=8mm\hbox{\raise -4mm\hbox{\epsfbox{link05.eps}}}}
\def\propagarr{\epsfxsize=12mm\hbox{\raise -1mm\hbox{\epsfbox{link06.eps}}}}
\Title{\vbox{\hbox{}\hbox{RUNHETC-99-37}\hbox{{\tt math-ph/9910010}}}}
{{\vbox {
\vskip-10mm
\centerline{Some Matrix Integrals related to Knots and Links}
}}}
\medskip
\centerline{P. Zinn-Justin}\medskip
\centerline{\it Department of Physics and Astronomy, Rutgers University,} 
\centerline{\it Piscataway, NJ 08854-8019, USA}

\vskip .2in

\noindent 
The study of a certain class of matrix
integrals can be motivated by their interpretation as counting
objects of knot theory such as alternating prime links, tangles or knots.
The simplest such model is studied in detail
and allows to rederive recent results
of Sundberg and \Th. The second non-trivial example turns out to be
essentially the so-called $ABAB$ model, though in this case the analysis
has not yet been carried out completely. 
Further generalizations are discussed.
This is a review of work done (in part) in collaboration with
J.-B.~Zuber. 

\bigskip
\vskip45mm

\noindent{to appear in the proceedings of the MSRI 1999 semester
on Random Matrices}
\Date{9/99}
%
\newsec{Introduction}
Using random matrices to count combinatorial objects is
not a new idea. It stems from the pioneering work
of \BIPZ, which showed how the perturbative expansion
of a simple non-gaussian matrix integral led, using
standard Feynmann diagram techniques, to the counting
of discretized surfaces. It has resulted in many applications:
from the physical side, it allowed to define a discretized version
of 2D quantum gravity \DFGZJ\ and to study various statistical models
on random lattices \refs{\Kaz,\KoS}. From the mathematical side, let us cite
the Kontsevitch integral \refs{\Kon,\Wit,\IZb},
and 
the counting of meanders and foldings \refs{\MAK,\DF}.

Here we shall try to apply this idea to the field of knot theory.
Our basic aim will be to count knots or related objects.
In the next sections,
we shall define these objects; then will follow a brief overview
of matrix models and how they can be related to knots. We shall
then explain the counting of alternating links (following \PZJZ)
in section 4;
study a generalized model (the $ABAB$ model \KP) in section 5, which
will lead us to digress and consider summations over Young tableaux;
and finally, discuss further generalizations in section 6.

\newsec{Knots, links and tangles}
Let us recall basic definitions of knot theory (from a physicist's
point of view; the reader is referred
to the literature for more precise definitions).
A {\it knot} is a smooth circle embedded in $\IR^3$.
A {\it link} is a collection of intertwined knots.
Both kinds of objects are considered up to homeomorphisms of $\IR^3$.
Roughly speaking, 
a {\it tangle} is a knotted structure from which four strings 
emerge.

In the $19^{\rm th}$ century, Tait introduced the idea
to represent such objects by their projection on the plane, 
with under/over-crossings at each double point and with {\it minimal number 
of such crossings} (fig.~\mincross).
\fig\mincross{A reduced diagram with 6 crossings.}{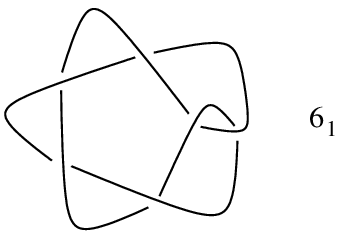}
We shall now consider such {\it reduced} diagrams. 
To a given knot, there corresponds
a finite number of (but not necessarily just one) reduced diagrams.
We shall come back later to
the problem of different reduced diagrams
which correspond to the same knot (or link, or tangle).

To avoid redundancies, we can concentrate on {\it prime}
links and tangles, whose diagrams cannot be
decomposed as a connected sum of components (Fig.~\nprime).
\fig\nprime{A non-prime diagram.}{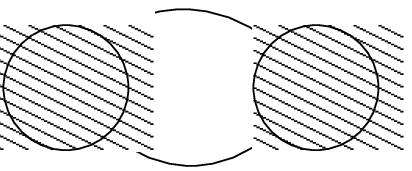}

A diagram is called {\it alternating} if one meets alternatively 
under- and over-crossings as one travels along each loop.\foot{Even though
it may not seem obvious, there are knots
that cannot be drawn in an alternating way -- starting with eight crossings.}
From now on, we shall concentrate on alternating diagrams only, since they are easier
to count. There are two reasons for that.

The first reason is that there is a simpler way to characterize if two reduced
alternating 
diagrams correspond to the same knot or link (than the general Reidemeister
theorem \ref\REID{K. Reidemeister, {\sl Knotentheorie}, Springer (1932).}).
%
Indeed, a major result conjectured by Tait and proven by Menasco and \Th~\MTh\
is that two alternating reduced knot or link diagrams represent 
the same object if
and only if they are related by a sequence of moves acting on tangles 
called ``flypes'' (see Fig.~\flyp).
\epsfxsize=8cm\fig\flyp{The flype of a tangle.}{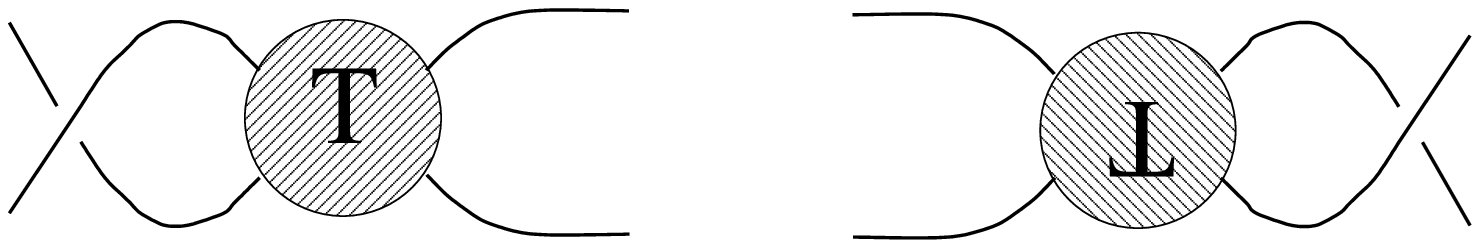}

The second reason is that there is a
correspondance between alternating diagrams and planar diagrams 
(see for example \KAUF),
which will be explained now as we discuss matrix integrals.

\newsec{Matrix integrals}
Let us now start from a completely different angle and consider
the following matrix integral:
\eqn\onemm{Z^{(N)}(g)=\int \d M\, \e{ N\, \tr \left(-\oh  M^2
+{g\over 4} M^4\right)}}
$M$ is a $N\times N$ hermitean matrix;
$g$ is a real parameter, which should be chosen negative to make
the integral convergent.

As an application of Wick's theorem,
the perturbative expansion of $Z^{(N)}$ in powers of $g$ can be made using
the following Feynman rules: one should count all diagrams
made out of vertices
$\vertexdbl=gN \delta_{qi}\delta_{jk}$
and propagators 
$\bra M_{ij} M^\dagger_{k\ell}\ket_0
=\propagdbl ={1\over N}\delta_{i\ell}\delta_{jk}$.
Due to the double lines, these diagrams form so-called fat graphs
which can be identified with triangulated surfaces.
Each diagram has a weight 
$$(gN)^{\rm V} N^{-\rm E} N^{\rm F}\, {1\over\scriptstyle{\rm symmetry\ factor}}$$
where V, E, F are
the number of vertices, edges, faces of the triangulated surface
(the factor $N^{\rm F}$ comes from the summation over internal indices).
The symmetry factor (i.e.\
the order of the automorphism group of the diagram) is of little
importance to us and we shall not discuss it any further.
Note that the power of $N$ is simply $N^\chi$ where $\chi$ is
the Euler--Poincar\'e characteristic of the triangulated surface.
If we take the logarithm, which amounts to considering only
connected surfaces, then we have the following genus expansion:
$$\log Z^{(N)}(g)= \sum_{h=0}^\infty F_h(g) N^{2-2h}$$
where $F_h$ is the sum over surfaces of genus $h$.
In particular, if we consider the large $N$ limit, we see that
$$F(g)=\lim_{N\to\infty}{\log Z^{(N)}(g)\over N^2}=\sum_{\hbox{\rm planar\
graphs}} {g^{\rm V}
\over\scriptstyle{\rm symmetry\ factor}}$$
is the sum over connected ``planar'' diagrams (i.e.\ with spherical topology).
$F(g)$ is the quantity we are interested in. The formal power series
$F(g)=\sum_p f_p g^p$ turns out to have, as is well-known, a finite radius
of convergence (which allows to analytically continue it to positive
values of $g$, as will be explained later). The position and nature
of the closest singularity $g_c$ determines the asymptotics of $f_p$ as
$p\to\infty$ i.e.\ of the number of planar diagrams with large
numbers of vertices.

In order to connect with knot theory, we take any planar diagram
and do the following: starting from an arbitrary crossing,
we decide it is a crossing of two strings (again there is an arbitrary
choice of which is under/over-crossing). Once the first choice is made,
we simply follow the strings and form alternating sequences of
under- and over-crossings. The remarkable fact is that this can
be done consistently (Fig.~\corresp).
If we identify two alternating
diagrams obtained from one another by inverting undercrossings
and overcrossings, then there is a one-to-one correspondence
between planar diagrams and alternating link diagrams. So the
function $F(g)$ also counts alternating link diagrams with
a given number of crossings.
\epsfxsize=10cm\fig\corresp{A planar diagram 
and the corresponding alternating link diagram.}{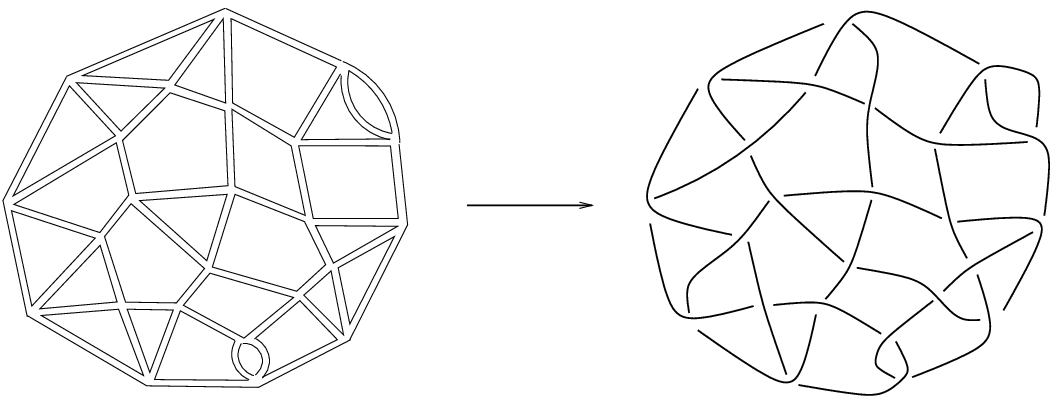}

A more detailed discussion of
the properties of the resulting link diagrams will be made in the
next section. For now, we shall address the question of the number of
connected components of the link (as a 3-dimension object). 
Indeed, there is no reason for the diagram
to represent a simple knot, and not several intertwined knots. In order
to distinguish them, we introduce a more general model,
which we shall call the intersecting loops $O(n)$ model.
If $n$ is a positive 
integer, then consider the following multi-matrix integral:
\eqn\mmm{
Z^{(N)}(n,g)=\int\! \prod_{a=1}^n \d M_a
\, \e{N\,\tr\left(-{1\over 2} \sum_{a=1}^n M_a^2+{g\over 4} \sum_{a,b=1}^n
M_a M_b M_a M_b\right)}}
and the corresponding free energy
\eqn\mmmb{F(n,g)=\lim_{N\to\infty}{\log Z^{(N)}(n,g)\over N^2}}
This model has an $O(n)$-invariance where the $M_a$ behave as a vector
under $O(n)$.
Its Feynman rules are a bit more complicated since
we should draw the diagrams with $n$ different colors. The colors
``cross'' each other at vertices just like strings in links (Fig.~\colors).
\fig\colors{A planar diagram with 2 colors.}{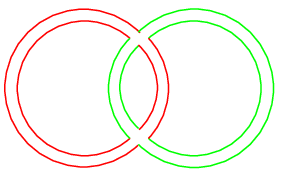}
So what we have done is allow each loop in the link to have $n$ different
colors. This is in itself an interesting generalization of the original
counting problem. Indeed, we can write:
\eqn\mmmc{F(n,g)=\sum_{k=1}^\infty F^k(g) n^k}
where $F^k(g)$ is the sum over alternating link diagrams with
exactly $k$ intertwined knots. We see that the links are weighted
differently according to their number of connected components.

But there is more. The expression \mmmc\ is an expansion of
$F(n,g)$ as a function of $n$ around $0$;
it provides a definition of $F(n,g)$ for non-integer values of $n$.
In particular, we have the following formal expression
$$F^1(g)=\dd{F(n,g)}{n}_{\displaystyle | n=0}$$
for the sum over alternating knots (this is the classical replica trick).
Therefore, if one computed $F(n,g)$ for arbitrary (non-integer) values
of $n$, one would have access to the generating function of the number of
alternating knots.
Of course, it might seem difficult to solve our model for all $n$;
we shall discuss this again in the conclusion.

\newsec{The one matrix model and the counting of links}
Let us now come back to the one-matrix model and show how one can
derive explicit formulae for the counting of prime alternating links.
We recall the partition function
\eqn\onemmb{Z^{(N)}(g)=\int \d M \e{N\, \tr \left( -\oh M^2
+{g\over 4}M^4\right)}}
and the corresponding free energy
\eqn\onemmc{F(g)=\lim_{N\to\infty} {1\over N^2} \log Z^{(N)}(g)
}
We also define the correlation functions:
\eqn\onemmd{
G_{2n}(g)=\lim_{N\to\infty}\bra {1\over N}\tr M^{2n}\ket}
Whereas the perturbative expansion of $F(g)$ generates closed
digrams (and therefore alternating links),
the $G_{2n}(g)$ count diagrams with $2n$ external legs.
In particular, we shall be interested later in
$\Gamma(g)=G_4(g)-2G_2(g)^2$ which counts connected diagrams with
$4$ legs, i.e.\ alternating tangles.

There are various methods to compute all these quantities. We shall
briefly recall the simplest one: the saddle point method. 

\subsec{Saddle point method for the one-matrix model}
We start
from Eq.~\onemmb\ and notice that the action and measure are $U(N)$-invariant;
therefore we can go over to the eigenvalues $\lambda_i$ of $M$:
\eqn\onemme{Z(g)=\int\prod_{i=1}^N \d\lambda_i 
\Delta[\lambda_i]^2
\,\e{N\, \sum_{i=1}^N \left( -\oh  \lambda_i^2
+{g\over 4}\lambda_i^4\right)}}
up to an overall constant factor. Here $\Delta[\cdot]$ is the Van der Monde
determinant. In Eq.~\onemme, the action is of order $N^2$ while there
are only $N$ variables of integration. Therefore, in the large $N$ limit,
a saddle point analysis applies. It is easy to see that as $N\to\infty$,
the eigenvalues $\lambda_i$ form a continous saddle point 
density $\rho(\lambda)$
whose support is an interval $[-2a,2a]$. In order to solve the
saddle point equations, it is convenient to introduce the resolvent
\eqn\onemmf{
\eqalign{
\omega(\lambda)&=\lim_{N\to\infty}
\left<{1\over N}\tr{1\over\lambda -M}\right>\cr
&=\int_{-2a}^{2a} \d\lambda'{\rho(\lambda')\over\lambda-\lambda'}}}
Then the saddle point equations read:
\eqn\onemmg{\omega(\lambda+i0)+\omega(\lambda-i0)
-\lambda+g\lambda^3=0\qquad\lambda\in[-2a,2a]}
This is a simple Riemann--Hilbert problem which can be solved:
\eqn\onemmh{\omega(\lambda)={1\over2}\lambda-{1\over2}g\lambda^3
-\left(-{1\over2}g\lambda^2+{1\over2}-ga^2\right
)\sqrt{\lambda^2-4a^2}}
with $a^2={1\over6}{1-\sqrt{1-12g}\over g}$.
What we have found is the generalized semi-circle law (for $g=0$ we recover
the usual GUE). 
Since $\omega(\lambda)$ is a generating function
of the $G_{2n}$, we can extract
$$G_2(g)=
\hbox{\raise-0.33cm\hbox{\epsfbox{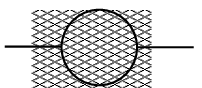}}}\,
={1\over 3} a^2 (4-a^2)$$
$$\Gamma(g)=G_4(g)-2G_2(g)^2=
\hbox{\raise-0.65cm\hbox{\epsfbox{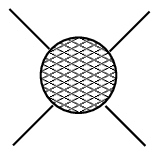}}}\,
=a^4 (a^2-1)(2a^2-5)$$
Also, we find
$$F(g)={1\over2}\log a^2 - {1\over 24} (a^2-1)(9-a^2)$$

Note that all these expressions can now
be analytically continued
to $g>0$ all the way to the singularity $g_c=1/12$. This has a simple
interpretation: changing the sign of $g>0$ corresponds to making
the potential in which the eigenvalues lie unstable; however, there
is still a local minimum at the origin and since the large $N$ limit
is a {\it classical limit}, the eigenvalues cannot quantum tunnel
to the unstable region and therefore remain in the valley (Fig.~\valley).
However, as
$g$ reaches its critical value $g_c$, the eigenvalues
begin overflowing, which causes the singularity.
\fig\valley{The potential for $g>0$; analytic continuation
is possible as long as the eigenvalues stay trapped inside
the central valley.}{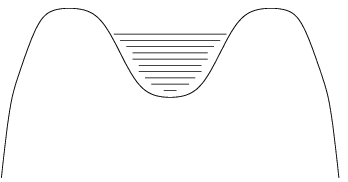}

Of course, $F(g)$ is not yet the counting function of 
prime alternating links. There are two separate problems to resolve:

\item{1)} Are the diagrams {\it reduced} (i.e.\ do they have
minimal crossing number)? Do they
correspond to {\it prime} links?

\item{2)} What about the flype equivalence? One should count
only once different diagrams which are flype-equivalent.

We shall address them now.

\subsec{Primality and Minimality}
The diagrams obtained from the matrix model can have
``nugatory'' crossings 
or ``non-prime'' parts (Fig.~\nonprim).
\fig\nonprim{First terms in the perturbative expansion of the 2-point function.}{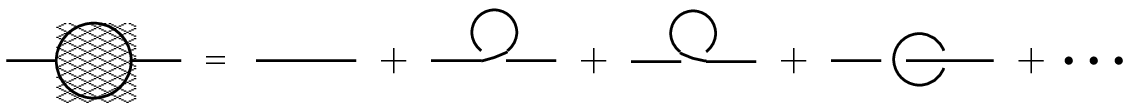}
Note however that all
these unwanted features appear as part of the two-point function.
Therefore, in order to remove them, we must simply set the
two-point function equal to 1! This is achieved by introducing
an additional parameter $t$ in the action:
\eqn\onemmpm{ Z^{(N)}(t,g)=\int \d M\, \e{ N\, 
\tr \left(-{t\over2}  M^2
+{g\over 4} M^4\right)}}
Of course, $t$ can be absorbed in a rescaling of $M$, so the
model is essentially unchanged. However we can now ask that
$t$ be chosen as a function of $g$ such that
\eqn\tg{G_2(t(g),g)=1}
We can solve this equation;
the auxiliary function $a(g)$ introduced earlier is now the solution of a third
degree equation
\eqn\thideg{27 g= (a^2-1)(4-a^2)^2 }
equal to $1$ when $g=0$; and $t(g)$ is given by
\eqn\thidegb{ t(g)={1\over 3} a^2(g) (4-a^2(g))}
The function $\Gamma(g):=\Gamma(t(g),g)$ is then the
counting function for reduced alternating tangle diagrams.
Similarly, $F(g)$ defined by ${\d\over\d g}F(g)={1\over4}G_4(t(g),g)$
counts alternating link diagrams.
We find in particular
that the singularity of $F(g)=\sum_p f_p g^p$
(given by the equation $g_c/t^2(g_c)=1/12$)
has moved to $g_c=4/27$; taking into account the power of the singularity,
we find that the rate of growth of the number of alternating
diagrams with $p$ crossings is
\eqn\asy{
f_p\buildrel p\to\infty\over\sim 
{\rm const}\ 6.75^p\, p^{-{7\over 2}}}
A similar result was found in \Tutte.

\subsec{Flype equivalence}
The more serious problem we have to resolve is that we are
not really counting links: we are counting diagrams, and links
are flype-equivalence classes of diagrams. Here we shall follow \STh.

Let us take a closer look at the action of a flype (Fig.~\flyp). The key remark
is that it acts on tangles (i.e.\ four-point functions), but more
precisely on {\it two-particle reducible} (2PR) tangles.
This leads naturally to the idea of introducing {\it skeleton diagrams}:
a general connected tangle can be created
by putting {\it two-particle-irreducible} (2PI) diagrams in the ``slots''
of a {\it fully two-particle-reducible} skeleton diagram.
We then expect that the 2PR skeleton will be modified by the
flype-equivalence, whereas the
2PI pieces (or more precisely the corresponding
skeletons, see below) will be unaffected.
Let $\Gamma(g)= G_4(g)-2G_2(g)^2$ be the counting function of
connected tangles and $D(g)$ of
2PI tangles; then $\Gamma\{ D\}$, that is the power series obtained
by composing $\Gamma(g)$ and the inverse of $D(g)$,
is the counting function of fully 2PR skeleton diagrams (Fig.~\without).
\fig\without{The whole set of diagrams $\Gamma$ built out of the
2PI diagrams $D$.}{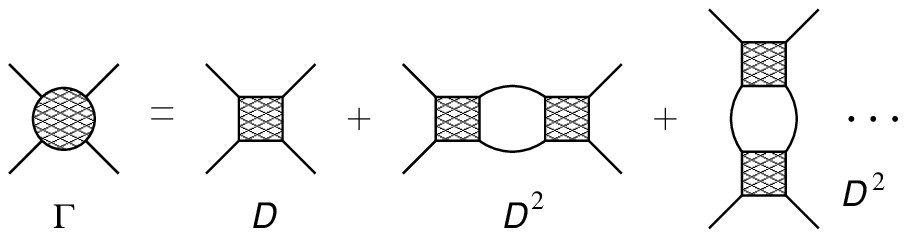}
It easy to see from general combinatorial arguments that $D(g)=\Gamma(g)
{1-\Gamma(g)\over 1+\Gamma(g)}$ and therefore
\eqn\eqone{\Gamma\{ D\}={1\over 2}\left[1-D-\sqrt{(1-D)^2-4D}\right]}

Inversely if $D(g)=g+\zeta(g)$, then $\zeta[\Gamma]$ is
the counting function of {\it fully 2PI} skeletons diagrams (Fig.~\zet).
\fig\zet{The set of 2PI diagrams built out of
general diagrams $\Gamma$ (plus the single crossing $g$).}{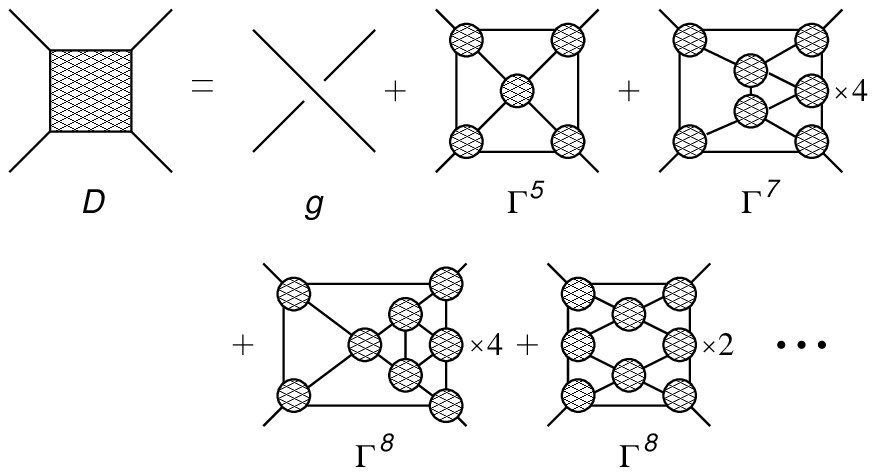}
From the solution of the one-matrix model, one obtains
\eqn\eqtwo{\zeta[\Gamma]=-{2\over 1+\Gamma}+2-\Gamma-{1\over2}
{1\over(\Gamma+2)^3}\left[1+10\Gamma-2\Gamma^2-(1-4\Gamma)^{3/2}\right]
}

As we mentioned earlier, after taking into account the flyping equivalence,
Eq.~\eqone\ will be modified, but not Eq.~\eqtwo. To see how it works,
let us show how the counting of Fig.~\without\ is redone (Fig.~\with).
\fig\with{Taking into account the flyping equivalence forces us
to distinguish simple crossings from non-trivial 2PI diagrams (marked
with a circle). There is only one term $g\zeta$ because the other
term is obtained by a flype.}{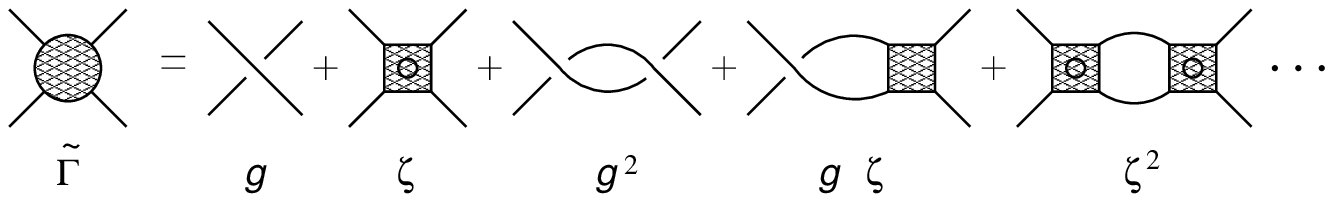}

More generally, 
we can redo the simple combinatorics to find the generating function
of the 2PR skeletons, but this time taking into account the flype
equivalence. We find
\eqn\eqonep{\tilde{\Gamma}\{g,\zeta\}=
{1\over 2}\left[ (1+g-\zeta) -\sqrt{(1-g+\zeta)^2-8 \zeta-8{g^2\over 1-g}}\right
]}
This is to be combined with the (unaltered) matrix model data
\eqn\eqtwop{\zeta[\Gamma]=-{2\over 1+\Gamma}+2-\Gamma-{1\over2}
{1\over(\Gamma+2)^3}\left[1+10\Gamma-2\Gamma^2-(1-4\Gamma)^{3/2}\right]
}
In practice, this means that
$\tilde{\Gamma}(g)$ is given by an implicit equation:
\eqn\eqimp{
\tilde{\Gamma}(g)=\tilde{\Gamma}\{g,\zeta[\tilde{\Gamma}(g)]\}
}
which can be reduced to a fifth degree equation. From the generating function
of tangles $\tilde{\Gamma}(g)$ we
can go back to the generating function of closed diagrams $\tilde{F}(g)$; we find
in particular that the singularity has been displaced again, so that if
$\tilde{F}(g)=\sum_{p=0}^\infty \tilde{f}_p g^p$, then
\eqn\finasy{\tilde{f}_p \buildrel p\to\infty\over\sim 
{\rm const}\ \left(
{101+\sqrt{21001}\over 40}
\right)^p\, p^{-7/2}}
where
$(101+\sqrt{21001})/40= 6.14793\ldots$ 
This result was first obtained in \STh.

\newsec{The $ABAB$ model and character expansion}
We shall now inspect the $n=2$ case of the general $O(n)$ model \mmm. There are various
reasons that this model is of particular interest, and we shall discover some of them
along the way. Let us rewrite the partition function
\eqn\abab{
Z^{(N)}(2,g)=\int\! \d A\d B
\, \e{N\,\tr\left(-{1\over 2} (A^2+B^2)+{g\over 4}(A^4+B^4)+{g\over 2} (AB)^2\right)}}
We see that we could introduce two coupling constants $\alpha$ and $\beta$:
\eqn\ababc{
Z^{(N)}_{ABAB}(\alpha,\beta)=\int\! \d A\d B
\, \e{N\,\tr\left(-{1\over 2} (A^2+B^2)+{\alpha\over 4}(A^4+B^4)+{\beta\over 2} (AB)^2\right)}}
(which amounts to introducing ``interaction'' between the two colors of loops).
For $\alpha=\beta=g$ we recover the $O(2)$ model. The more general model
with $\alpha$ and $\beta$ arbitrary is not
necessary for the original counting problem, but since it turns out that we can solve it
equally easily, we shall keep the two coupling constants.
Note that when $\alpha\ne\beta$ the $O(2)$ symmetry of the model is broken. This is
even more apparent if we make the change of variables $X={1\over\sqrt{2}}(A+iB)$,
$X^\dagger={1\over\sqrt{2}}(A-iB)$:
\eqn\abcabc{
Z^{(N)}_{8v}(b,c,d)=\int \d X\d X^\dagger\,
\e{ N\,\tr\left(-XX^\dagger
+bX^2 X^{\dagger 2}
+{c\over2}(XX^\dagger)^2
+{d\over4}(X^4+X^{\dagger 4})
\right)
}}
with $b=(\alpha+\beta)/2$ and $c=d=(\alpha-\beta)/2$. We recognize in \abcabc\ the partition
function of the {\it 8-vertex model} on random dynamical 
lattices.\foot{Or, more
precisely, a two-parameter slice of it, since $c$ and $d$ 
are not independent.} A configuration of the model is 
defined by a quadr-angulated surface with arrows on the edges
of the graph, such that each of the vertices displays one of the eight allowed
configurations, which are weighted with the 3 constants $b$, $c$, $d$.
For $\alpha=\beta$ the $U(1)$-breaking term 
$X^4+X^{\dagger4}$ vanishes and we recover the 
{\it 6-vertex model}.\foot{In this case note that the arrows 
``cross'' each other just like knots in links,
so that we manifestly recover our link model with the 2 orientations
of the loops playing the same role as the 2 colors.}

We shall now show how to solve the model in the planar limit, i.e.\ 
compute the large $N$ free energy.

\subsec{Character expansion}
All known matrix model solutions are (more or less implicitly) based on the fact
that we can reduce the number of degrees of freedom from $N^2$ to $N$. Usually the
$N$ remaining degrees of freedom are the eigenvalues of the matrices. Unfortunately,
from Eq.~\ababc\ one cannot go directly to the eigenvalues of $A$ and $B$: we
do not know how to integrate over the relative angle between $A$ and $B$.\foot{Only one
integral of this type is known exactly, the Harish Chandra--Itzykson--Zuber integral 
\refs{\HC,\IZ},
but it does not apply here.} Therefore, instead of working directly with \ababc,
we expand the troublesome part $\exp(N{\beta\over2}\tr(AB)^2))$ in {\it characters} of $GL(N)$.
Recalling that all class-functions can be expanded on the basis of characters, we write:
\eqn\charexp{
\e{N{\beta\over2}\tr(AB)^2}=\sum_{\{ h\}} c_{\{ h\}}
\chi_{\{ h\}}(AB)}
where $\chi_{\{ h\}}(AB)$ is the character taken at $AB$ and $\{ h\}$ is the set of
shifted highest weights $h_i=m_i+N-i$ ($m_i$ highest weights),
$h_1>h_2>\cdots>h_n\ge 0$,
which parametrize the $GL(N)$ analytic irreducible representation.
The coefficients of the expansion $c_{\{ h\}}$ can be determined explicitly:
\eqn\coeffexp{c_{\{ h\}}=(N\beta/2)^{\# h/2}
{\Delta(h^{\rm even}/2)
\Delta((h^{\rm odd}-1)/2)
\over\prod_i \lfloor h_i/2\rfloor !}}
in terms of the set $\{h^{\rm even}\}$ and $\{h^{\rm odd}\}$ of even and
odd $h_i$.
The advantage of characters is that they satisfy orthogonality relations,
so that we can now integrate over
the relative angle between $A$ and $B$:
\eqn\intUN{\int_{U(N)} \d\Omega\, \chi_{\{ h\}}(A\Omega B\Omega^\dagger)=
{\chi_{\{ h\}}(A)\chi_{\{ h\}}(B)\over\chi_{\{ h\}}(1)}}
where the dimension $\chi_{\{ h\}}(1)$ is up to an overall constant
the Van der Monde determinant $\Delta[h_i]$.

Once Eqs.~\charexp\ and \intUN\ are inserted into \ababc,
we see that the integrand only depends on the eigenvalues of $A$ and
$B$:
\eqn\charexpb{
Z^{(N)}_{ABAB}(\alpha,\beta)=\sum_{\{ h\}}
{c_{\{ h\}}\over\Delta[h_i]}
\left[\int\prod_{i=1}^N \d\lambda_i
\,\e{N\sum_{i=1}^N
\left(-{1\over2}\lambda_i^2+{\alpha\over 4}\lambda_i^4\right)}\,
\Delta[\lambda_i] \det_{i,j}[\lambda_i^{h_j}]
\right]^2}

The key observation here is that we still have an action of order
$N^2$, but we have $N$ highest weights $h_i$ and $N$ eigenvalues
$\lambda_i$; therefore a saddle point analysis applies again.

\subsec{Saddle point on Young tableaux}
The notion of a saddle point on Young tableaux first appeared
in \VERSH\ in the context of the asymptotics of the Plancherel measure.
It was rediscovered independently in the solution of large $N$
2D Yang--Mills \DK, and was used
to deal with character expansions in \KKSW.
In the present calculation, the novelty is that we have to deal with a
{\it double} saddle point equation on both eigenvalues and
shifted highest weights (i.e.\ shape of the Young tableau) \KP.

The idea here is to find an appropriate scaling ansatz for the
shape of the dominant Young tableau in the large $N$ limit.
We find that the highest weights $h_i$ scale as $N$
(the Young tableaux become large both horizontally and vertically),
so we can define a continuous density of rescaled $h_i/N$:
$$\rho(h)={1\over N}\sum_{i=1}^N \delta(h-h_i/N)$$
and the corresponding resolvent
$$H(h)=\int \d h'{\rho(h')\over h-h'}$$
We also have a density of eigenvalues $\rho(\lambda)$ and
the resolvent $\omega(\lambda)$.

The saddle point equations now read (we use ``slashed'' functions 
defined by: $\Hsl(h):={1\over2}(H(h+i0)+H(h-i0))$
and similarly for the other functions):
\eqn\dblspe{
\left\{\eqalign{
-\lambda+\alpha\lambda^3+\omegasl(\lambda)+\hsl(\lambda)/\lambda=0
\qquad &\lambda\in[-\lambda_0,+\lambda_0]\cr
\Lsl(h) - {\Hsl(h)\over 2} = {1\over 2} \log(h/\beta)
\qquad &h\in[h_1,h_2]}
\right.}
with $L(h)=\log \lambda^2(h)$. The new unknown functions $h(\lambda)$
and $\lambda(h)$ appear when taking the logarithmic derivative
of $\det_{i,j}[\lambda_i^{h_j}]$; this type of functions was analyzed
in \PZJ, where it was shown that $\lambda(h)$ and $h(\lambda)$ are
{\it functional inverses} of each other. Therefore, we have two
saddle point equations which are connected by a functional inversion
relation.
This connection allows to solve them; skipping the details, one
can show that one has a well-defined Riemann--Hilbert problem
for the
auxiliary function $D(h):=2L(h)-H(h)-3\log h +\log(h-h_1)$,
whose solution can be expressed in terms of $\Theta$ functions
in an appropriate elliptic parametrization $y(h)$:
$$
D(h)
=\log{h-h_1\over -\alpha h^2} 
- {\log(\beta/\alpha)\over K} y(h) 
+ 2 \log{\Theta_2(x_0-y(h))\over\Theta_2(x_0+y(h))}
$$

\subsec{Phase diagram and discussion}
In the same way that the one-matrix model displayed a singularity
at $g_c=1/12$, here the free energy and the various correlation functions
have a line of singularities in the
$(\alpha,\beta)$ plane, which is shown on Fig.~\phdiag.
\epsfxsize=11cm
\fig\phdiag{Phase diagram of the $ABAB$ model. The dashed line
is the $\alpha=\beta$ line, the curves are equipotentials
of the elliptic nome.}{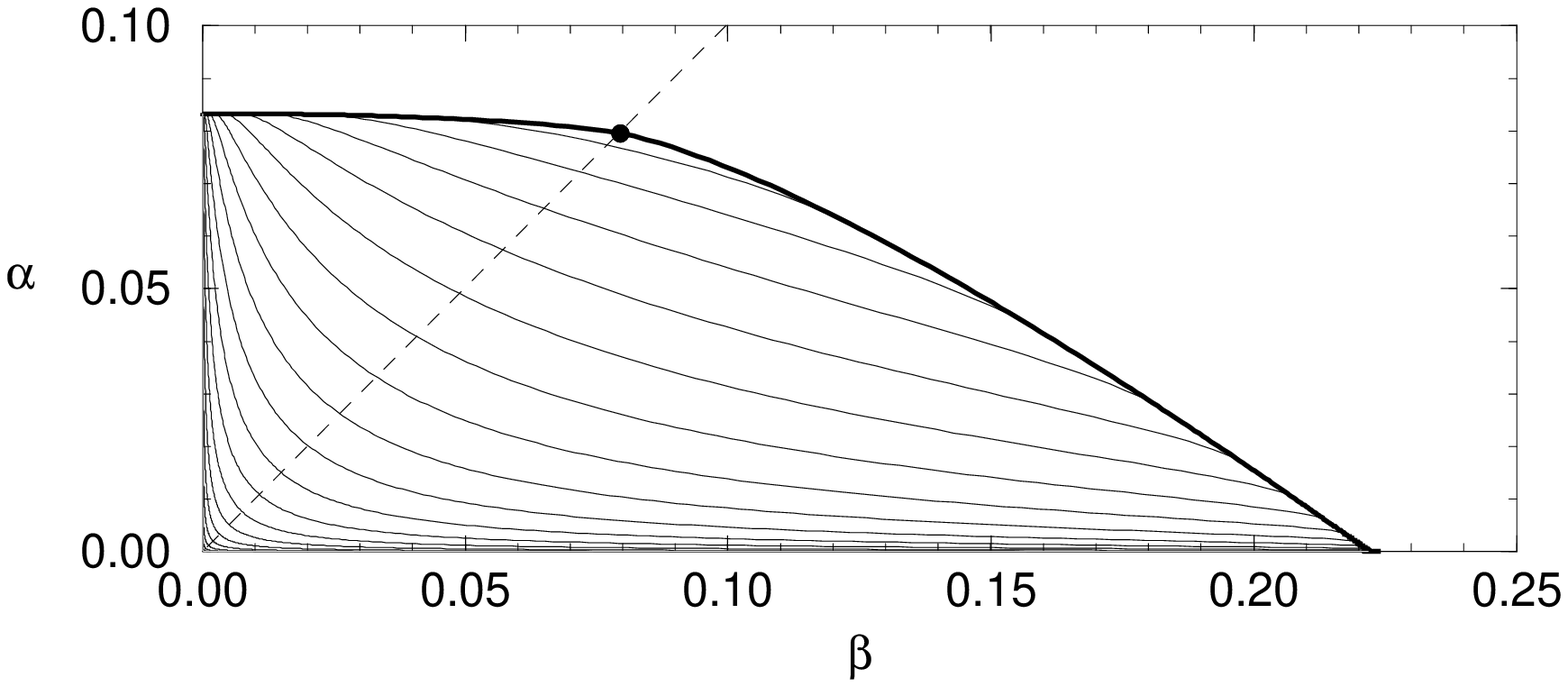}
We recognize at $\alpha=1/12$, $\beta=0$ the usual singularity
of the one-matrix model. In fact, one can show that everywhere
on the critical line except at the critical point $\alpha_c=\beta_c=
{1\over 4\pi}$, the critical behavior is the same as the one of the
one-matrix model (``pure gravity'' behavior).
This implies the following asymptotics:
at fixed slope $s=\beta/\alpha\ne 1$, if the free
energy $F(\alpha,\beta=s\alpha)=\sum_p f_p(s) \alpha^p$ then 
$$f_p(s) \buildrel p\to\infty\over\sim 
{\rm const}\ \alpha_c(s)^{-p}\, p^{-7/2}$$

However, the point $\alpha_c=\beta_c={1\over 4\pi}$, that is the 6-vertex
model point, is very special: this is the point where the elliptic
functions degenerate into trigonometric functions, which implies
logarithmic corrections:
$$f_p(1) \buildrel p\to\infty\over
\sim {\rm const}\ (4\pi)^p\, p^{-3} \log p$$
This is characteristic of a $c=1$ conformal field theory coupled
to gravity.

\subsec{Application to reduced alternating diagrams}
We should remember that the $\alpha=\beta$ line is of special
interest to us, since it is the 
intersecting loops $O(2)$ model (solving a certain counting problem
for alternating links) we started from.
In order to carry out the program that we have applied to the
one-matrix model we should next address the two issues of 
primality/minimality and of the flype equivalence.
We shall only consider the first issue; a dicussion of the flype
equivalence in the general $O(n)$ case will be made in the next section,
and the corresponding calculation for $n=2$ will appear
in a future publication \ref\FUT{P. Zinn-Justin and J.-B. Zuber,
work in progress.}.

Again, we introduce an additional parameter $t$ in the action:
\eqn\twommpm{
Z^{(N)}(2,t,g)=\int\! \d A\d B
\, \e{N\,\tr\left(-{t\over 2} (A^2+B^2)+{g\over 4}(A^4+B^4)+{g\over 2} (AB)^2\right)}}
and we impose that the 2-point function $G_2(t,g)
=\lim_{N\to\infty}\bra {1\over N}\tr A^2\ket
=\lim_{N\to\infty}\bra {1\over N}\tr B^2\ket$ satisfies
\eqn\tgb{
G_2(t(g),g)=1
}
Obvious scaling properties imply that $G_2(t,g)={1\over t} G_2(1,g/t^2)$, and
the formula for $G_2(1,g)$ can be found in appendix A of \KP\ in terms of complete elliptic
integrals. This gives an equation for $t(g)$, which can in principle be solved (at least
to an arbitrary order in perturbation theory).

In order to go further, we notice that at the singularity we must have
$g_c/t(g_c)^2={1\over 4\pi}$; from \KP\ we extract $G_2(1,1/(4\pi))={\pi\over2}(4-\pi)$,
and therefore using \tgb, $t(g_c)={\pi\over2}(4-\pi)$, which finally yields
\eqn\gcrit{g_c={\pi\over 16}(\pi-4)^2}
We conclude that the number $f_p$ of reduced alternating link
diagrams with $2$ colors and
$p$ crossings has the following asymptotics:
$$f_p \buildrel p\to\infty\over
\sim {\rm const}\ \left({16\over\pi(\pi-4)^2}\right)^p\, p^{-3} \log p$$
where the number $1/g_c=6.91167\ldots$ is slightly larger than the value $6.75$ obtained
for only one color.

\newsec{Further generalizations and prospects}
We have already written a fairly general model, the
intersecting loops $O(n)$ model (Eq.~\mmm) which should contain in principle all
information on the counting of alternating links and knots. We shall now show how
this model is in fact not sufficient for our purposes.

Indeed, one should remember that the $O(n)$ model given above counts
alternating link diagrams, and not alternating links. For the latter, one
should address
the problem of the flype equivalence. We have seen in the $n=1$ case
(section 4) that we needed to do a little surgery on the four-point functions. 
One can convince oneself that
what it amounts to,
in more physical terms, 
is a {\it finite renormalization} which results
in the appearance of quartic counterterms in the action. 
Generically these counterterms
will have the most general form compatible with the symmetry.
In our case, we find that there are two independent $O(n)$-symmetric tetravalent
vertices, of the form $M_a M_b M_a M_b$ and $M_a M_a M_b M_b$.
This results in a generalized $O(n)$ model:
\eqn\mmmgen{
Z^{(N)}(n,g,h)=\int\! \prod_{a=1}^n \d M_a
\, \e{N\,\tr\left(-{1\over 2} \sum_{a=1}^n M_a^2
+{g\over 4} \sum_{a,b=1}^n (M_a M_b)^2
+{h\over 2} \sum_{a,b=1}^n M_a^2 M_b^2
\right)}}
where $h$ will be given as a function of $g$ by appropriate combinatorial relations
of the same form as those of section 4.

At the moment, the solution of this general model is unknown. Let us note however
that for $h=0$, this model is simply the usual (non-intersecting loops) $O(n)$ model
which has been completely solved \KoS. It is tempting to speculate
that there is no phase transition in the $(g,h)$ plane as one moves away from
the $h=0$ line.\foot{This is certainly true in the $n=2$ case,
as the study of \DAL\ shows, and the exact solution \PZJb\ confirms.}
Then, one can make predictions on {\it universal quantities} such
as critical exponents of the model. For example, the number $\tilde{f}_p(n)$
of prime alternating links with $n$ colors would have the asymptotics
$$\tilde{f}_p(n) \buildrel?\over\sim {\rm const}(n)\ 
b(n)^p\, p^{-2-1/\nu}\qquad n=-2\cos(\pi\nu),\ 0< \nu< 1$$
In particular the number $\tilde{f}_p$ of prime alternating knots would satisfy
$$\tilde{f}_{p}\buildrel?\over\sim
{\rm const}\ b(0)^p\, p^{-4}$$
One interesting question is to determine the non-universal constant
$b(0)$. This, of course, requires to really solve the $n=0$ model
(or more precisely to study the $n\to 0$ limit).
An alternative option is to note
that this model can be recast as a supersymmetric $Osp(2n|2n)$ model,
the simplest ($n=1$) being a supersymmetrized version of the $O(2)$ model
considered earlier; using bosonic and fermionic
complex matrices $X$ and $\Psi$, it can be written as:
\eqn\susy{\eqalign{
Z^{(N)}&(g)=\int \d X\d X^\dagger\,\d\Psi\d\Psi^\dagger\cr
&\e{ N\,\tr\left(-XX^\dagger-\Psi\Psi^\dagger
+g(X X^\dagger X^\dagger X
+\Psi\Psi^\dagger\Psi^\dagger\Psi
+\Psi X^\dagger\Psi^\dagger X
+X^\dagger\Psi X\Psi^\dagger
)\right)}\cr
}}
Due to supersymmetry, this partition function is equal to $1$,
but non-supersymmetric correlation functions such as $\left<{1\over N}
\tr(XX^\dagger)^n\right>$ are non-trivial and should contain the
desired information. Whether this model is solvable or not is an open
question.
\vskip1cm
\centerline{\bf Acknowledgements} 
I would like to thank J.B.~Zuber for various discussions and with
whom most of this work was done.
I also want to thank Prof. D.~Eisenbud
for the hospitality of the MSRI, and the organizers
of this semester on Random Matrices, P.~Bleher and  A.~Its, for inviting me
and giving me the opportunity to give seminars.


\listrefs
\bye

\bye